
\input phyzzx
\voffset = -0.4in \footline={\ifnum\pageno=1 \nulline
\else\newfootline \fi} \def\nulline{{\hfill}}
\def\newfootline{\advance\pageno by -1\hss\tenrm\folio\hss}
\rightline {March 1995}
\rightline {QMW-TH-95-29.}
\title {SPECIAL GEOMETRY\break AND TWISTED MODULI  IN  ORBIFOLD
THEORIES \break WITH CONTINUOUS WILSON LINES.}
\author{W. A. Sabra*, S. Thomas** and N. Vanegas**}
\address {* Department of Physics,\break Royal Holloway and
Bedford New College,\break University of London,\break Egham,
Surrey, U.K.} \address {** Department of Physics,\break Queen
Mary and Westfield College,\break University of London,\break
Mile End Road , London E1 4NS, U.K.} \abstract {Target space
duality symmetries,  which acts on  K\"ahler and continuous
Wilson line  moduli, of a ${\bf Z}_N$ ($N\not=2$) 2-dimensional
subspace of  the moduli space of orbifold compactification are
modified to include twisted moduli. These spaces described by
the cosets $SU(n,1)\over SU(n)\times U(1)$  are $special$
K\"ahler,  a fact which is exploited in deriving the extension
of tree level duality transformation to include
higher orders of the twisted moduli.  Also, restrictions on these
higher order terms are derived.}
\endpage
\REF\one{L. Dixon, J. A. Harvey, C. Vafa and E. Witten, {\it
Nucl. Phys.} {\bf B261} (1985) 678; {\bf B274} (1986) 285.}
\REF\two{ A. Font, L. E. Ib\'a\~nez, F. Quevedo and A. Sierra,
{\it Nucl. Phys.} {\bf B331} (1991) 421.} \REF\three{S.  Ferrara,
D.  L\"ust, and S.  Theisen,     {\it Phys. Lett.} {\bf B233}
(1989) 147; S.  Ferrara, D.  L\"ust,  A. Shapere, and S.
Theisen,{\it Phys. Lett.} {\bf B225} (1989) 363 ;  M. Cveti\v c,
A.  Font, L. E.  Ib\'{a}\~{n}ez,  D. L\"ust and  F.  Quevedo,
   {\it Nucl. Phys.} {\bf B361} (1991) 194; S.  Ferrara, C.
Kounnas, D.  Lust and F.  Zwirner,    {\it Nucl. Phys.} {\bf
B365} (1991) 431.} \REF\four{ L. Dixon, D. Friedan, E. Martinec
and S. H. Shenker, {\it Nucl. Phys.} {\bf B282} (1987) 13; S.
Hamidi and C. Vafa,  {\em Nucl. Phys.} {\bf B279} (1987) 465.}
\REF\five{L. Dixon, V. Kaplunovsky and J. Louis, {\it Nucl.
Phys.} {\bf B329} (1990) 27.} \REF\six{M. Cveti\v c, J. Louis and
B. Ovrut, {\it Phys. Lett.} {\bf B206} (1988) 229;  M. Cveti\v
c, J. Molera and B. Ovrut, {\em Phys. Rev.}{\bf D40} (1989)
1140.} \REF\cec{ S. Cecotti, S. Ferrara and L. Girardello, {\it
Nucl. Phys.} {\bf B308} (1989) 436;  {\it Phys. Lett.}  {\bf
B213}  (1988) 443.}
\REF\seven{G. L. Cardoso, D. L\"{u}st and T. Mohaupt,  {\it
Nucl. Phys.} {\bf B432} (1994) 68; M. Cveti\v c, B. Ovrut and W.
A. Sabra,   {\it Phys. Lett.} {\bf B351} (1995) 173; P. Mayr and
S. Stieberger, hep-th 9412196.}
\REF\eight{E. Witten, {\it Phys. Lett.} {\bf B155} (1985) 151;
S. Ferrara, C. Kounnas and M. Porrati, {\it Phys. Lett.} {\bf
B181} (1986) 263; S. Ferrara, L. Girardello, C. Kounnas and M.
Porrati, {\it Phys. Lett.} {\bf  B192} (1987) 368; {\it Phys.
Lett.} {\bf B194}  (1987) 358; H. P. Nilles. {\it Phys. Lett.}
{\bf B180} (1986) 240; C. P. Burgess, A. Font and F. Quevedo,
{\it Nucl. Phys.} {\bf B272} (1986) 661; U. Ellwanger and M. G.
Shmidt, {\it Nucl. Phys.} {\bf B294 }(1987) 445; J. Ellis, C.
Gomez and D. V. Nanopoulos, {\it Phys. Lett.} {\bf B171}  (1986)
203.}
\REF\nine{E. Cremmer and A. Van Proeyen, {\it Class. and Quantum
Grav.} {\bf 2} (1985) 445; E. Cremmer, C. Kounnas, A. Van
Proeyen,  J. P. Derendinger, S. Ferrara, B. de Wit and L.
Girardello, {\it Nucl. Phys.} {\bf B250} (1985) 385; B. de Wit
and A. Van Proeyen, {\it Nucl. Phys.} {\bf B245} (1984) 89; A.
Strominger, {\it Commun. Math. Phys.} {\bf 133}(1990) 163; L.
Castellani, R. D'Auria and S. Ferrara, {\it Phys. Lett.} {\bf
B241} (1990) 57 and {\it Class. and Quantum Grav.} {\bf 1}(1990)
1767; R. D'Auria, S. Ferrara and P. Fr\`e, {\it Nucl. Phys.} {\bf
B359} (1991) {705}.}
\REF\ten{S. Ferrara, D. L\"ust and S. Theisen,  {\it Phys. Lett.}
{\bf B242}, 39 (1990).} \REF\eleven{L. E. Ib\'{a}\~{n}ez, H. P.
Nilles and F. Quevedo,  {\it Phys. Lett.} {\bf B192} (1987) 332;
L. E. Ib\'{a}\~{n}ez, J. Mas, H. P.  Nilles and F. Quevedo, {\it
Nucl. Phys.} {\bf B301} (1988) 157.} \REF\su{E. Cremmer, S.
Ferrara, L. Giraradello and A. Van Proeyen, {\it Nucl. Phys.}
{\bf B212} (1983) 413.} 
Heterotic string theories compactified on orbifolds  are of
phenomenological importance as they  give rise to an $N=1$
space-time supersymmetric semi-realistic four dimensional
quantum field theories [\one, \two].  A phenomenologically
appealing  feature of  string compactifications lies in the fact
that the physical couplings  in the low-energy effective action
are  dependent on the moduli fields which parametrize the shape
and size of the orbifold and possible continuous Wilson lines.
Also, such an action has a novel symmetry  known as target space
duality.  This is a discrete symmetry of the moduli space which
leaves the underlying conformal field theory invariant.  This
symmetry restricts the low energy effective action and connects
it to the theory of modular forms [\three].  The low-energy
action is an $N=1$ supergravity coupled to Yang-Mills and matter
fields.  If terms with up to  two derivatives in the bosonic
fields are included, the theory is then defined in terms of the
K\"ahler potential $K$ which encodes the kinetic terms for the
massless fields,  the superpotential $W$ containing the Yukawa
couplings and the $f-$function whose real part, at the tree
level, determines the gauge couplings [\su].  In fact the
lagrangian of the theory depends  on $K$ and $W$ via the target
space duality invariant combination  $${\cal G}=K+log\vert
W\vert^2.\eqn\sidon$$
The method of calculating the superpotential of the low-energy
effective action directly  from the underlying conformal field
theory was given in [\four]. Moreover, the untwisted moduli
dependence of the  tree level  K\" ahler potential has been
addressed  in the literature by several methods [\five-\eight].

It is our purpose in this letter to derive the explicit
dependence of the  K\"{a}hler potential on the full moduli space,
untwisted and twisted, of a 2-dimensional subspace of an orbifold
with  continuous Wilson lines [\eleven]. We will consider the
cases where the moduli space is given by the $special$ K\"{a}hler
manifold $SU(n,1)\over SU(n)\otimes U(1)$ [\nine].
This coset is parametrized by one K\"ahler modulus and $n-1$
complex continuous Wilson lines of a plane where the twist has
a complex eigenvalue \footnote*{this means that the twist is not
${\bf Z}_2$} [\seven].  The real part of the K\"ahler or toroidal
modlus describes the size of the 2-dimensional space  and the imaginary
part describes a possible internal axion field (antisymmetric
tensor). The Wilson lines are homotopically non-trivial flat
gauge connections.

In the process of deriving the K\"ahler potential, the higher
order corrections, in terms of the twisted moduli, to the target
space  duality symmetry of the underlying conformal field theory
are derived. Moreover, conditions on these higher order terms are
also determined. For simplicity we will concentrate on the case
with one complex Wilson line, $i.e.,$  the moduli space
$SU(2,1)\over SU(2)\otimes U(1)$,  and determine the full
K\"ahler potential of the moduli space in the presence of a
generic twisted modulus. Generalization to more than one Wilson
line or twisted modulus is straightforward.  The fact that the
moduli spaces $SU(n,1)\over SU(n)\otimes U(1)$ are special
K\"ahler facilitates the calculation  of the higher order duality
transformations  for the moduli and their associated K\"ahler
potential as has been demonstrated in [\ten] for the special
K\"ahler coset $\Big[{SU(1,1)\over U(1)}\Big]^3$.  This can be
explained as follows.  For special K\"ahler manifolds [\nine],
the K\"ahler potential can be expressed in terms of a holomorphic
function of the moduli, if we  denote the moduli by $\phi^i$ then
$$K=-log Y;\qquad  Y=\sum_i(\phi^i+\bar\phi^i)({ F}_{\phi^i}+\bar
{F}_{\bar\phi^i})-2({ F}+\bar  F).\eqn\sp$$
In terms of  the homogeneous coordinates $x^I, I=0,\cdots, j,$
where the  physical moduli are expressed by the $special$
coordinates, $\phi^i={\textstyle x^i\over\textstyle  i x^0} $
(with $i=1,\cdots j$), \sp\  is expressed in the form
$$K=-log\Big({{\bar {\cal F}}_Ix^I+{{\cal F}}_I{\bar x}^I\over
x^0{\bar {x}}^0}\Big) =-log\Big[{-2i\over x^0{\bar
{x}}^0}\pmatrix{x^I\cr {1\over2}i{\cal F}_J }^\dagger
\pmatrix{0&\delta^L_{\  I}\cr -\delta^J_{\  K}&0}\pmatrix{x^K\cr
{1\over2}i{\cal F}_L }\Big],\eqn\kp$$ where ${\cal F}(x)=-(x^0)^2
F(\phi).$ From \kp\ it is clear that the K\"ahler potential is
invariant up to a K\"ahler transformation under
$$\eqalign{x^I&\rightarrow U^I_{\ J}x^J+{1\over2}iV^{IJ}{\cal
F}_{J},\cr {1\over2}i{\cal F}_I&\rightarrow
W_{IJ}x^J+{1\over2}iZ_I^{\ J}{\cal F}_{J},}\eqn\sym$$ with
$$S=\pmatrix{U&V\cr W&Z},\qquad S^\dagger\eta S=\eta,\qquad
\eta=\pmatrix{0&{\bf 1}\cr -{\bf 1}&0}\eqn\ramzi$$ where ${\bf
1}$ is the $j$-dimensional identity. The transformations in
\ramzi\ act as holomorphic field redefinitions provided that
$$\eqalign{&W_{IJ}x^J+{1\over2}iZ_I^{\ J}{\partial {\cal
F}(x)\over\partial x^I}= {1\over2}i{\partial \tilde {\cal
F}(y)\over\partial y^I},\cr  &y^I= U^I_{\
J}x^J+{1\over2}iV^{IJ}{\partial {\cal F}(x)\over\partial
x^{J}}.}\eqn\s$$ The  integrability is guaranteed if $U, V, W$
and $Z$ are real and satisfy [\ten] $$U^tW-W^tU=0\quad
V^tZ-Z^tV=0\quad U^tZ-W^tV= I.\eqn\re$$ These transformatons
define  duality transformations if ${\cal F}={\tilde {\cal F}}.$

The importance of this formalism lies in the fact that the
duality transformations act linearly and are field independent.
Therefore, the lowest order duality transformations fix $S$
completely.  The second equation in \s\ determine the modified
duality transformations of the moduli, and  corrections to the
tree level  transformations comes from the higher order terms in
the holomorphic function $\cal F$.  Moreover the first eq in \s\
restricts the higher order terms of the function $\cal F$ itself.
Finally the knowledge of the function $\cal F$ is sufficient to
determine the  Yukawa couplings of the theory [\ten].

We now get back to the study of $SU(2,1)\over SU(2)\otimes U(1)$
moduli space.  To lowest order in the twisted modulus ${\bf C}$,
the tree level K\"ahler potential  is given by
$$K=-log ({\bf T}+\bar {\bf T} -{k\over2}{\bf A}\bar {\bf A}-{\bf
C}\bar {\bf C}),\eqn\ka$$
where $\bf T$ is the K\"ahler complex modulus and $\bf A$ is a
complex Wilson line and $c$ is a model-dependent constant (for
${\bf Z}_3$ twist $k={\sqrt3}$).  We would like now to generalize
\ka\ to include higher orders of the expectation values of the
twisted modulus $\bf C$.  To be concrete we take the twist to be
${\bf Z}_3.$
A duality symmetry of the theory is given by the subgroup
$SL(2,Z)$ which acts on  on the moduli as follows
$${\bf T}\rightarrow {a{\bf T}-ib\over ic{\bf T}+d}, \quad {\bf
A}\rightarrow {{\bf A}\over ic{\bf T}+d},\quad {\bf C}\rightarrow
{{\bf C}\over ic{\bf T}+d}\qquad ad-bc=1.\eqn\sl$$ This acts on
\ka\ by a K\"ahler transformation $K\rightarrow K+\ln\vert(ic{\bf
T}+d)\vert^2.$  However, it is more convenient to work with a
different basis for the moduli space.  Perform a change of
variable as follows: $${{\bf T}\over\sqrt3}={1-t\over 1+t},\qquad
{\bf A}={2{\cal A}\over (1+t)}, \qquad {\bf C}={{{\sqrt2}\cal
C}\over (1+t)},\eqn\trans$$ then in terms of the new variables
$t$, ${\cal A}$ and ${\cal C}$, the duality transformations \sl\
takes the form $$t\rightarrow {\tilde
t}={\alpha\over\gamma}={At+B\over Ct+D},\quad {\cal A}\rightarrow
\tilde{\cal A}={{\cal A}\over\gamma},  \quad {\cal C}\rightarrow
{{\cal C}\over \gamma},\eqn\transf$$ where
$$\eqalign{&A=D^*={1\over2}\Big((a+d)+i(b'-c')\Big),\quad
B=C^*={1\over2}\Big((d-a)+i(b'+c')\Big)\cr & b'={b\over\sqrt3},\
{c'=\sqrt 3}c.}\eqn\new$$  The K\"ahler potential in terms of the
new coordinates can be written in the form  $$K=-log (1-t\bar t
-{\cal A}\bar {\cal A}-{\cal C}\bar {\cal C}).\eqn\hell$$
The  holomorphic function which produces \hell\ via \sp\ is given
by  $$F(t,{\cal A}, {\cal C})=-{1\over4}(1+ t
^2+{\cal A}^2+{\cal C}^2)=-{1\over
4{(x^0)}^2}\Big((x^0)^2-\sum_i^3(x^i)^2\Big),\eqn\holo$$ where
$(t, {\cal A},{\cal C})=({\textstyle x^1\over \textstyle ix^0},
{\textstyle x^2\over\textstyle ix^0}, {\textstyle
x^3\over\textstyle  ix^0}).$   The duality transformation
\transf\ acts on the homogeneous coordinates as in \sym\ where the
matrices $U, V, W$ and $Z$ are given by
$$Z={1\over2}\pmatrix{(a+d)&(b'+c')&0&0\cr (b'+c')&(a+d)&0&0\cr
0&0&2&0\cr 0&0&0&2},\quad
U={1\over2}\pmatrix{(a+d)&-(b'+c')&0&0\cr -(b'+c')&(a+d)&0&0\cr
0&0&2&0\cr 0&0&0&2}$$ $$W=-{1\over8}\pmatrix{(c'-b')&(a-d)&0&0\cr
(a-d)&(c'-b')&0&0\cr 0&0&0&0\cr 0&0&0&0}, \quad
V=2\pmatrix{(c'-b')&(d-a)&0&0\cr (d-a)&(c'-b')&0&0\cr 0&0&0&0\cr
0&0&0&0}.\eqn\ma$$ If one expands  ${\cal F}$ in terms of higher
orders of the twisted moduli,
$${\cal F}={1\over4}\Big((x^0)^2-\sum_i^3(x^i)^2
-\sum_{n=0}^{\infty}f_n(t, {\cal A}){
{(x^3)}^{n+3}\over {(ix^0)}^{n+1}}\Big),\eqn\mff$$ then using
\kp\ the  K\"ahler potential is now given by
$$\eqalign{&K=-ln\Big[1-t\bar t-{\cal A}\bar {\cal A}-{\cal
C}\bar {\cal C}-\cr & {1\over4}\sum \Big([(t+\bar t){\partial
f_n\over\partial t}+({\cal A}+\bar{\cal A}){\partial
f_n\over\partial {\cal A}}+(n+1)f_n ]{\cal C}^{n+3}+ (n+3)f_n
\bar{\cal C}{\cal C}^{n+2}+c.c\Big)\Big].}\eqn\mkp$$  Moreover,
making use of \mff\ and the second equation in \s, the modified
duality transformations of the moduli  can be determined and are
given by  $$\eqalign{&t\rightarrow t'={\alpha-{1\over4}\sum\Big(
(d-a)[t{\textstyle\partial f_n\over\textstyle\partial t}+ {\cal
A}{\textstyle\partial f_n\over\textstyle\partial {\cal A}}+
(n+1)f_n]+i(c'-b'){\textstyle\partial f_n\over\textstyle\partial
t}\Big){\cal C}^{n+3}\over\Gamma},\cr &{\cal A}\rightarrow {\cal
A}'={{\cal A}\over\Gamma},\qquad {\cal C}\rightarrow {\cal
C}'={{\cal C}\over\Gamma},\cr &\Gamma=\gamma-{1\over4}\sum\Big(
i(c'-b')[t{\partial f_n\over\partial t}+{\cal A}{\partial
f_n\over\partial{\cal A}}+(n+1)f_n] -(d-a){\partial
f_n\over\partial t}\Big){\cal C}^{n+3}\cr \  \  \
&=\gamma+\sum_nk_n {\cal C}^{n+3}.}\eqn\low$$ Clearly under these
transformations  the K\"ahler potential transforms as
$K\rightarrow K+\ln\Gamma+\ln\bar\Gamma.$ Using \mff\ and the
first equation \s\ we obtain
$$\eqalign{&\gamma-{1\over4}\sum\Big( (a+d)[t{\partial
f_n\over\partial t}+{\cal A}{\partial f_n\over\partial{\cal
A}}+(n+1)f_n]+i(b'+c') {\partial f_n\over\partial t}\Big){\cal
C}^{n+3}\cr  &= \Gamma\Big\{1-{1\over2}\sum \Big(t'{\partial
f'_n\over\partial {t'}}+ {\cal A}'{\partial
f'_n\over\partial{{\cal A}'}}+(n+1)f'_n\Big){{\cal
C}'}^{n+3}\Big\},}$$ $$\eqalign{&\alpha-{1\over4}\sum\Big(
i(b'+c')[t{\partial f_n\over\partial t}+{\cal A}{\partial
f_n\over\partial{\cal A}}+(n+1)f_n]-(a+d) {\partial
f_n\over\partial t}  \Big){\cal C}^{n+3}\cr  &=
\Gamma\Big(t'+{1\over2}\sum {\partial f'_n\over\partial
{t'}}{{\cal C}'}^{n+3}\Big),}$$ $${\cal A}+{1\over2}\sum
{\partial f_n\over\partial{\cal A}}{\cal C}^{n+3}  =
\Gamma\Big({{\cal A}'}+{1\over2}\sum {\partial
f'_n\over\partial{{\cal A}'}}{{\cal C}'}^{n+3}\Big),$$ $${\cal
C}+{1\over2}\sum (n+3)f_n{\cal C}^{n+2}  = \Gamma({{\cal
C}'}+{1\over2}\sum (n+3) f'_n{{\cal C}'}^{n+2}\Big).\eqn\beirut$$
 Expanding $t'$ and ${\cal A}'$ around $\tilde t$ and $\tilde
{\cal A}$ given in equation \transf, we get $$\eqalign{
&t'=\tilde t+{1\over 4\gamma^2}\Big\{ \sum
\Big(i(c'-b')\alpha-(d-a)\gamma\Big)\Big(t{\partial
f_n\over\partial t}+{\cal A}{\partial f_n\over\partial{\cal
A}}+(n+1)f_n\Big) \cr &+\Big((a-d)\alpha -i(c'-b')
\gamma\Big){\partial f_n\over\partial t}\Big\}{\cal
C}^{n+3}\sum_{m=0}^{m=\infty}\Big({l\over\gamma}\Big)^m\cr &=
\tilde t+\sum_n (\Delta t)_n{\cal C}^{n+3},\cr &{\cal
A}'=\tilde{\cal A}+{{\cal A}l\over\gamma^2}
\sum_{m=0}^{m=\infty}\Big({l\over\gamma}\Big)^m  = \tilde {\cal
A}+\sum_n (\Delta {\cal A})_n{\cal C}^{n+3}.}\eqn\hellbound$$
where $$l={1\over4}\sum\Big(i(c'-b')[t{\partial f_n\over\partial
t}+{\cal A}{\partial f_n\over\partial{\cal
A}}+(n+1)f_n]+(a-d){\partial f_n\over\partial t}\Big) {\cal
C}^{n+3}.\eqn\pinhead$$ Using \beirut-\pinhead, one could
determine the conditions that the functions $f_n$ must satisfy
and for the first five terms one gets,
$$\eqalign{f_0(\tilde t,\tilde {\cal A})=&\gamma f_0(t, {\cal
A}),\cr  f_1(\tilde
t,\tilde {\cal A})=&\gamma^2f_1(t, {\cal A}),\cr  f_2(\tilde t,
\tilde{\cal A})=&\gamma^3f_2(t, {\cal A}),\cr f_3(\tilde t,
\tilde{\cal A})=&\gamma^4f_3-{\gamma^5\over2}{\partial
f_0\over\partial {\cal A}}(\Delta {\cal A}_0 +C{\cal A}\Delta
t_0) -{\gamma^5\over2}\Delta t_0C f_0 -{\gamma^6\over2}\Delta
t_0{\partial f_0\over\partial t} +{\gamma^3\over2}k_0 f_0,\cr
f_4(\tilde t, \tilde{\cal A})=&\gamma^5f_4+{3\over7}\gamma^4 k_1
f_0+{8\over7}k_0\gamma^4f_1 -{4\over7}\Delta
t_0\Big(2\gamma^6Cf_1+\gamma^7{\partial f_1\over\partial
t}+\gamma^6C{\cal A}{\partial f_1\over\partial {\cal A}}\Big)\cr
& -{4\over7}\Delta {\cal A}_0\gamma^6{\partial
f_1\over\partial{\cal A}} -{3\over7}\Delta
t_1\Big(\gamma^6Cf_0+\gamma^7{\partial f_0\over\partial
t}+\gamma^6C{\cal A}{\partial f_0\over\partial{\cal A}}\Big)
-{3\over7}\Delta{\cal A}_1\gamma^6{\partial f_0\over\partial
{\cal A}}.}\eqn\calor$$   The $SL(2, Z)$ duality symmetry is only
a subgroup of the full duality symmetry of the theory.
Next we employ another duality symmetry subgroup
of the theory in order to derive further constraints on the functions $f_n.$
A symmetry of the theory is given by \footnote*{In
terms of the old basis $\bf T$ and $\bf A$, this is given by
${\bf A}\rightarrow {\bf A}+p$, \  ${\bf T}\rightarrow {\bf T}+
{\sqrt3\over2}p {\bf A}+{\sqrt 3\over4}p^2$, where $p$ is a multiple of
${\sqrt2}$.}
$$\eqalign{t&\rightarrow \hat t={\hat A\over\hat
B}={t-{\textstyle p\over\textstyle2} {\cal A}-{\textstyle
p^2\over\textstyle8}(1+t)\over  1+{\textstyle
p\over\textstyle2}{\cal A}+{\textstyle
p^2\over\textstyle8}(1+t)},\quad  {\cal A}\rightarrow \hat{\cal
A}={\hat C\over\hat B}={{\cal A}+{\textstyle
p\over\textstyle2}(1+t)\over  1+{\textstyle
p\over\textstyle2}{\cal A}+{\textstyle
p^2\over\textstyle8}(1+t)},\cr &{\cal C}\rightarrow {{\cal
C}\over \hat B}.}\eqn\pa$$ This symmetry  acts on the homogeneous
coordinates as given in \sym\ where the matrices $U, V, W$ and
$Z$ are given by $$U=Z= \pmatrix{1+{\textstyle
p^2\over\textstyle8}&0&0&0\cr 0&1-{\textstyle
p^2\textstyle\over\textstyle8}&-{\textstyle
p\over\textstyle2}&0\cr  0&{\textstyle p\over\textstyle2}&1&0\cr
0&0&0&1},\quad V=16W=\pmatrix{0&{\textstyle
p^2\over\textstyle2}&2p&0\cr -{\textstyle
p^2\over\textstyle2}&0&0&0\cr 2p&0&0&0\cr 0&0&0&0}.\eqn\hia$$
Using \hia\ and the second eq in \s, the modification of the
duality transformations \pa\ can be calculated. These modified
transformations are given by
$$\eqalign{&t\rightarrow
t'={{t-{\textstyle p\textstyle\over\textstyle2} {\cal
A}-{\textstyle p^2\textstyle\over\textstyle8}(1+t)}+{\textstyle
p^2\textstyle \over\textstyle16}\sum\Big(t{\textstyle\partial
{f_n}\textstyle\over\textstyle\partial t}+ {\cal
A}{\textstyle\partial {f_n}\over\textstyle\partial{\cal
A}}+(n+1)f_n\Big){\cal C}^{n+3}\over \Lambda},\cr &{\cal
A}\rightarrow {\cal A}'={ {\cal A}+{\textstyle
p\over\textstyle2}(1+t)-{\textstyle
p\over\textstyle4}\sum\Big(t{\textstyle\partial
{f_n}\over\textstyle\partial t}+{\cal A} {\textstyle\partial
f_n\over\textstyle\partial{\cal A}}+(n+1)f_n\Big){\cal
C}^{n+3}\over\Lambda},\cr &{\cal C}\rightarrow {\cal C}'={{\cal
C}\over\Lambda},\cr &\Lambda=1+{p\over2}{\cal
A}+{p^2\over8}(1+t)+\sum\Big({p^2\over16}{\partial f_n\over
\partial t}+{p\over4} {\partial f_n\over \partial  {\cal
A}}\Big){\cal C}^{n+3}.}\eqn\sh$$  Expanding $t'$ and ${\cal A}'$
as given in \sh\ around $\hat t$ and $\hat {\cal A}$ we obtain
$$\eqalign{t'=& {\hat A+X_1\over\Lambda}={\hat A+X_1\over \hat
B-X_2}= {\hat t}+{1\over \hat B^2} (\hat AX_2+\hat
BX_1)\sum_{m=0}^\infty ({X_2\over\hat B})^m= {\hat t}+\sum_n
(\delta t)_n {\cal C}^{n+3},\cr {\cal A}'=&{\hat C+X_3\over \hat
B-X_2}={\hat {\cal A}}+{1\over \hat B^2} (\hat CX_2+\hat
BX_3)\sum_{m=0}^\infty ({X_2\over\hat B})^m= {\hat {\cal
A}}+\sum_n (\delta{\cal A})_n {\cal C}^{n+3},\cr {\cal C}'=&{
{\cal C}\over \hat B}-{{\cal C}\over \hat B^2} \sum_n s_n{\cal
C}^{n+3}, \cr \Lambda=&\hat B+\sum_n s_n{\cal
C}^{n+3}.}\eqn\michel$$
Using the first eq in \s\ and \michel\
the following conditions on the first five of the functions $f_n$
are obtained
$$\eqalign{&f_0(\hat t, \hat {\cal A})=B f_0(t,
{\cal A}),\cr  &f_1(\hat t, \hat {\cal A})=B^2f_1(t, {\cal
A}),\cr  &f_2(\hat t, \hat{\cal A})=B^3f_2(t, {\cal A}),\cr
&f_3(\hat t,\hat {\cal A})=B^4f_3 -{B^6\over2}\delta {\cal
A}_0\Big[({p\over2}f_0+B{\partial f_0\over\partial {\cal
A}})(1+{p\over2}{\cal A})+({{p^2\over8}}f_0+B{\partial
f_0\over\partial {t}}) {p\over2}(1+t)\Big]\cr &-{B^6\over2}\delta
t_0\Big[({p\over2}f_0+B{\partial f_0\over\partial {\cal
A}})(-{p\over2}-{p^2\over8}{\cal A})+
({p^2\over8}f_0+B{\partial f_0\over\partial
{t}})(1-{p^2\over8}(1+t) )\Big]\cr &+{B^3\over2}s_0f_0,\cr
&f_4(\hat t, \hat{\cal A})=B^5f_4+{3\over7}B^4 s_1
f_0+{8\over7}s_0B^4f_1\cr  &-{3\over7}B^7\delta {\cal
A}_1\Big[({p\over2}f_0+B{\partial f_0\over\partial {\cal
A}})(1+{p\over2}{\cal A})+({{p^2\over8}}f_0+B{\partial
f_0\over\partial {t}}) {p\over2}(1+t)\Big]\cr
&-{3\over7}{B^7}\delta t_1\Big[({p\over2}f_0+B{\partial
f_0\over\partial {\cal A}})(-{p\over2}-{p^2\over8}{\cal
A})+({{p^2\over8}}f_0+B{\partial f_0\over\partial {t}})(1-
{p^2\over8}(1+t))\Big]\cr & -{4\over7}{B^7}\delta
t_0\Big[(pf_1+B{\partial f_1\over\partial {\cal A}})(-{p\over
2}-{p^2\over8}{\cal A})+({p^2\over4} f_1+B{\partial
f_1\over\partial t})(1-{p^2\over8}(1+t))\Big]\cr
&-{4\over7}B^7\delta {\cal A}_0\Big[(pf_1+B{\partial
f_1\over\partial {\cal A}})(1+{p\over2}{\cal A})+({p^2\over4}
f_1+ B{\partial f_1\over\partial
{t}}){p\over2}(1+t)\Big].}\eqn\mal$$

In conclusion, by employing the methods  of special geometry,
duality symmetries of the coset space
$SU(2,1)\over SU(2)\otimes U(1)$, parametrized by the complex
K\"ahler moduli $\bf T$  and one complex Wilson moduli $\bf A$
of a ${\bf Z}_3$ two-dimensional plane of an orbifold
compactifaction, are realized as symplectic transformations
on the vector whose components are the homogeneous coordinates
$x^I$ and ${1\over2}i{\partial {\cal F}\over\partial x^I},$
where $\cal F$ is the holomorphic function describing the special
K\"ahler geometry. Using the fact that
such tansformations  are exact to all orders in the expansion of
$\cal F$ in the twisted moduli, constraints are derived on the
moduli dependent coefficient of the expansion. The choice of
these functions depends on the model under consideration. It
should be mentioned that only a subspace of the duality symmetry
of the theory have been implemented in deriving such constraints
and it would be interesting to derive a set of constraints which
are obtainable from the full duality symmetry group.  We hope to
report on this in a future publication.
\centerline{\bf ACKNOWLEDGEMENT}
W. S would like to thank M. Cveti\v c for many useful
discussions and D. L\"ust, A. Van Proeyen and S.
Stieberger for comments.
The work of W. S is supported by P.P.A.R.C, S. T
by the Royal Society and N. Vanegas by Colciencias and
Universidad de
Antioquia (Colombia) studentship.
\vfill\eject
\refout
\end